\documentclass[epj]{webofc}
\usepackage[varg]{txfonts}   
\usepackage{color}
\usepackage{epsfig}
\wocname{EPJ Web of Conferences}
\woctitle{New Frontiers in Physics 2014}

\begin{document}
\selectlanguage{english}
\title{Status of the  Thermal Model and Chemical Freeze-Out}
%
%

\author{Jean Cleymans\inst{1}\fnsep\thanks{\email{jean.cleymans@uct.ac.za}}
}

\institute{UCT-CERN Research Centre and Department of Physics\\
	University of Cape Town, Rondebosch 7701, South Africa
          }

\abstract{%
A brief review  is presented  of the status of the thermal model and chemical freeze-out 
in relativistic heavy-ion collisions. Some interewsting aspects at lower energies are emphasized.
}
\maketitle
\begin{figure}[htb]
\centerline{\epsfig{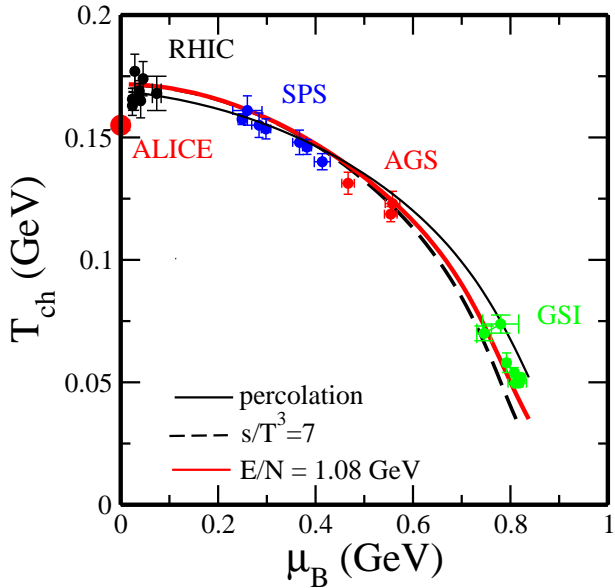}}
\caption{Chemical freeze-out temperature $T$ vs. the baryon chemical potential  at different beam 
energies together with curves corresponding to various models~\cite{wheaton}.}
\label{crete2014}
\end{figure}
\begin{figure}[htb]
\centerline{\epsfig{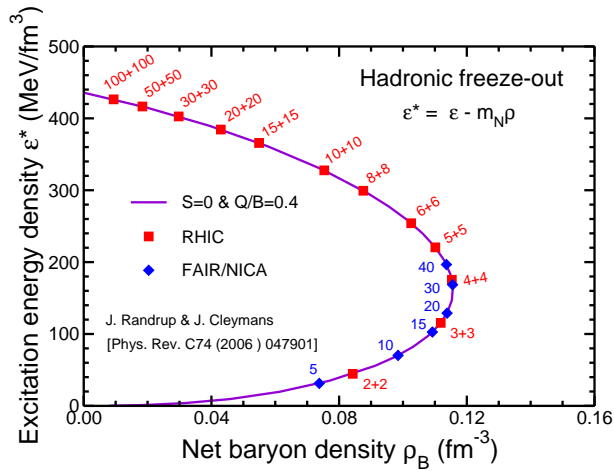}}
\caption{The hadronic freeze-out line in the $\rho_B-\epsilon^{\*}$ phase plane 
as obtained from the values of $\mu_B$ and $T$
 that have been extracted from the experimental data in \cite{wheaton}.
The calculation employs values of $\mu_Q$ and $\mu_S$ 
that ensure $\langle S\rangle=0$ and $\langle Q\rangle=0.4\langle B\rangle$
for each value of $\mu_B$.  
Also indicated are the beam energies (in GeV/N)
for which the particular freeze-out 
conditions are expected at either RHIC or FAIR or NICA. 
}
\label{randrup_figure}
\end{figure}
\begin{figure}
\centerline{\epsfig{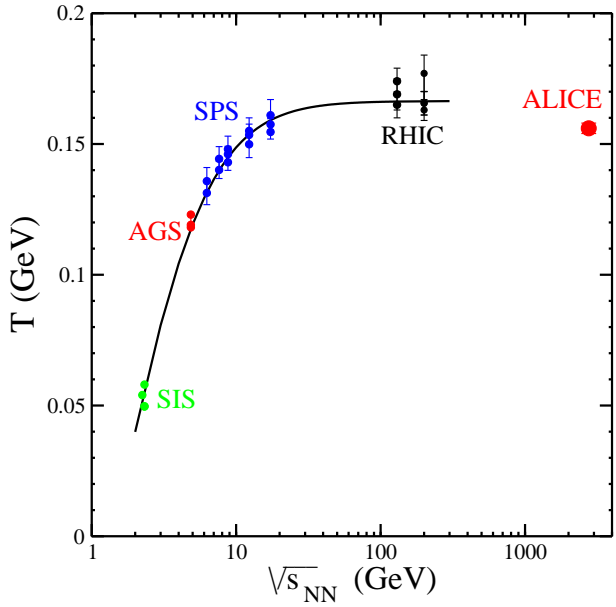}}
\caption{Chemical freeze-out temperature $T$ as a function of the beam energy.}
\label{tvse}
\end{figure}
\begin{figure}
\centerline{\epsfig{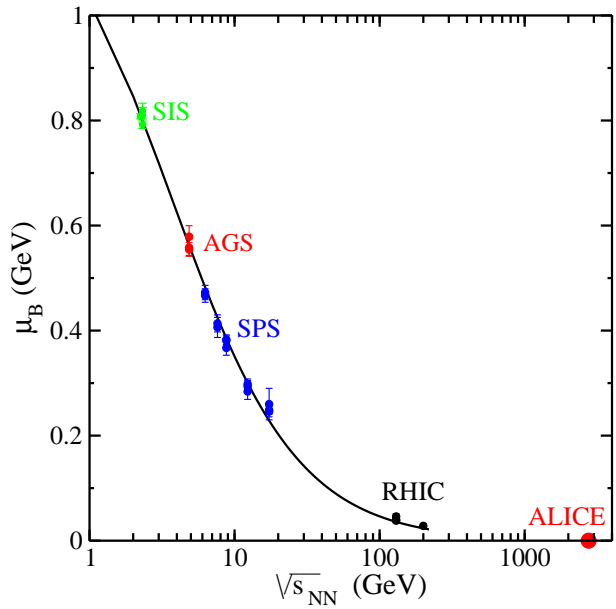}}
\caption{Chemical freeze-out baryon chemical potential $\mu_B$ as a function of the beam energy.}
\label{mubvse}
\end{figure}
\begin{figure}[htb]
\centerline{\epsfig{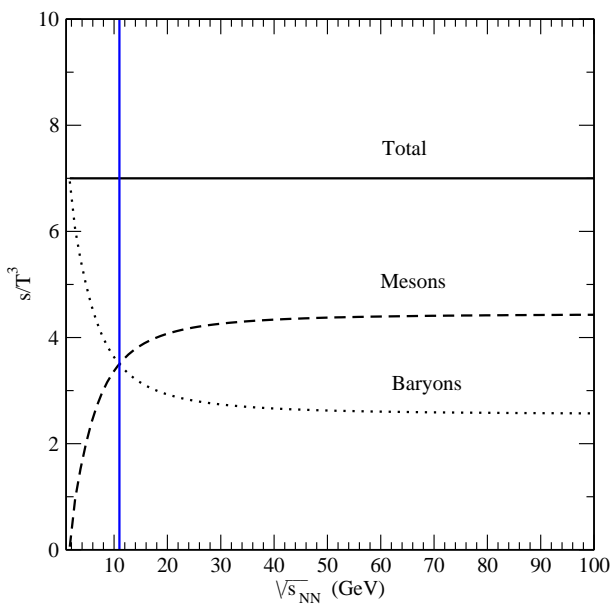}}
\caption{The $s/T^3$ ratio calculated in the thermal-statistical model along the constant value consistent with
chemical freeze-out. Also shown are the contributions from the mesons and the baryons.}
\label{sovert3}
\end{figure}
\begin{figure}[htb]
\centerline{\epsfig{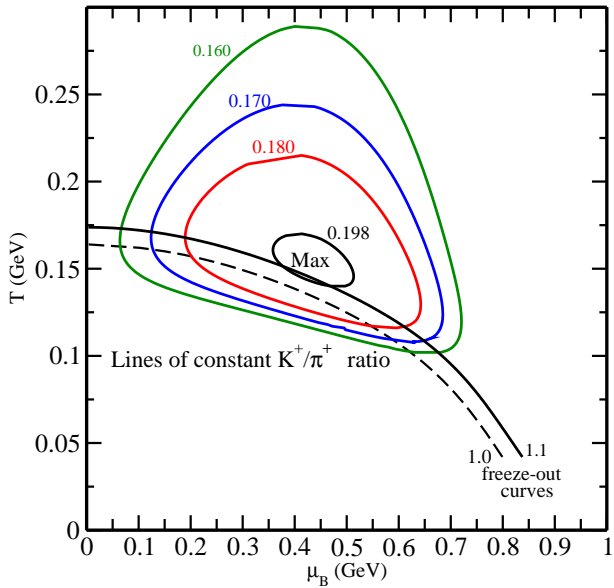}}
\caption{Lines of constant value of the $K^+/\pi^+$ ratio in the $T-\mu_B$ plane showing a clear maximum 
in this ratio close to the boundary given by the chemical freeze-out line.}
\label{kpluspiplus}
\end{figure}

%
Particle collisions at high energies produce large numbers of secondaries and it is natural to try
a statistical-thermal model to analyse these.  This type of analysis has a long 
and proud history~\cite{koppe,fermi,hagedorn}. \\
In  relativistic heavy ion collisions a new dimension is given to the model by the varying  baryon density.
The highly successful analysis of particle yields lead to the notion of chemical 
equilibrium which is by now a  well-established tool in the analysis of relativistic heavy ion 
collisions, see e.g.~\cite{cleymans-satz}. 
Before the start of RHIC, only three points were present on the $T-\mu_B$ plane, showing a clear increase of the 
chemical freeze-out temperature, $T$,
with increasing beam energy and an accompanying decrease of the baryon chemical  potential $\mu_B$~\cite{becattini,review}.
Substantial new knowledge became available in the following decade~\cite{pbm,manninen,wheaton} and now cover almost 
the complete $T-\mu_B$ curve as shown in Fig.~1. A last substantial gap
still exists in an energy region to be covered by the Beam Energy Scan at RHIC~\cite{shusu} and by FAIR and NICA.
The results obtained at ALICE have been presented comprehensively recently in~\cite{michele} with a 
chemical freeze-out temperature of 155 $\pm$ 2 MeV which is slightly lower than expected.
A comparison~\cite{wheaton} of three parameterizations is shown in Fig.~\ref{crete2014}.
The resulting freeze-out curve in the $T-\mu_B$ plane can also be drawn in the
energy density vs net baryon density plane as was first done in Ref.~\cite{randrup}. The
resulting curve is shown in Fig.~\ref{randrup_figure}.
This figure shows that the highest net baryon density will be reached in the beam energy covered by the RHIC/FAIR/NICA 
experiments.
In view of the success of  chemical freeze-out  in relativistic heavy ion collisions, 
much effort has gone into finding models that lead to a final state in chemical equilibrium, see e.g.  
curve~\cite{magas_satz,transition,biro,castorina}.
The corresponding dependence of the temperature and the chemical potential on beam energy is 
surprisingly smooth~\cite{wheaton} as shown in Figs.~\ref{tvse} and~\ref{mubvse}.
However, despite this smoothness in the thermal freeze-out parameters a 
roller-coaster is observed in several  particle ratios, e.g. the horn in the $K^+/\pi^+$ ratio and a similar
strong variation in the $\Lambda/\pi$ ratio~\cite{NA49}.
Again these strong variations are not observed in $p-p$ collisions.
Within the framework of thermal-satistical models this variation has been connected to a change from
a baryon domicated to a meson dominated hadron gas~\cite{transition}. This conclusion is based on the observation that the entropy density
divided by the temparature to the third power, $s/T^3$, is constant over the whole energy range. 
The change is shown in Fig.~\ref{sovert3}.
Lines of constant value for the $K^+/\pi^+$ ratio are shown in Fig.~\ref{kpluspiplus} where it can be seen that the 
absolute maximum in the thermal-statistical model hugs the chemical freeze-out line. The largest observed value
is just barely compatible with this maximum. 
In the thermal-statistical model
a rapid change is expected as the hadronic gas undergoes a
transition from a baryon-dominated  to a meson-dominated gas. The
transition occurs at a temperature $T$ = 151 MeV and baryon
chemical potential $\mu_B$ = 327 MeV corresponding to an incident
energy of $\sqrt{s_{NN}}$ = 11 GeV.  
Thus the strong variation seen in the particle ratios
corresponds  to a transition from a baryon-dominated to
a meson-dominated hadronic gas. This transition occurs at a
\begin{itemize}
\item temperature $T = $ 151 MeV, 
\item baryon chemical potential $\mu_B = $ 327 MeV, 
\item  energy $\sqrt{s_{NN}} = $ 11 GeV. 
\end{itemize}
In the
statistical model this transition leads to peaks in the
$\Lambda/\left<\pi\right>$,  $K^+/\pi^+$, $\Xi^-/\pi^+$ and
$\Omega^-/\pi^+$ ratios. However, the observed ratios are sharper than the ones 
calculated in thermal-statistical models.\\
We present a review of data obtained in relativistic heavy ion collisions and show that there
is a gap around 11 GeV where more and better precise measurements are needed.
The theoretical interpretation can only be clarified by covering this 
energy region.
In particular the strangeness content needs to be determined, data covering
the full phase space (4$\pi$) would be very helpful to determine the thermal parameters of 
a possible phase transition and the existence of a quarkyonic phase as has been discussed recently~\cite{mclerran}.
\section*{Acknowledgments}
The numerous contributions by H  Oeschler,  J. Randrup, K. Redlich, E. Suhonen 
and S. Wheaton are gratefully acknowledged.

\end{document}